\documentclass[aps,pra,showpacs,amsmath,amssymb]{revtex4}
\usepackage[usenames,dvipsnames]{color}
\usepackage[toc,page]{appendix}
\usepackage{epsf}
\usepackage{bm}
\usepackage{dcolumn}
\usepackage{latexsym}
\usepackage{amsmath}
\usepackage{amsfonts}
\usepackage{amssymb}
\usepackage{graphicx}
\usepackage{braket}
\graphicspath{}
\usepackage[active]{srcltx}
\newcommand{\be}{\begin{eqnarray}}
\newcommand{\ee}{\end{eqnarray}}

\definecolor{darkred}{rgb}{.8,0,0}

\definecolor{darkblue}{rgb}{0,0,.7}

\begin{document}
%
%%%%%%%%%%%%%%%%%%%%%%%%%%%%%%%%%%%%%%%%%%%%%%%%%%%%%%%%%%%%%%%%%%%%%%%%%%%%%%%%%%%%%%%
%%%%%%%%%%%%%%%%%%%%%%%%%%%%%%%%%%%%%%%%%%%%%%%%%%%%%%%%%%%%%%%%%%%%%%%%%%%%%%%%%%%%%%%
\title{Quantum-enhanced atomic gyroscope with tunable precision}
%%%%%%%%%%%%%%%%%%%%%%%%%%%%%%%%%%%%%%%%%%%%%%%%%%%%%%%%%%%%%%%%%%%%%%%%%%%%%%%%%%%%%%%
%
\author{J. P. Cooling}
%\email{J.Cooling@sussex.ac.uk}
%
%\affiliation{Department of Physics and Astronomy, University of Sussex, Brighton, BN1 9QH, United Kingdom\\}
%
%\author{L. M. Rico-Gutierrez}
%\affiliation{Universidad del SABES Celaya, Manuel Orozco y Berra 101, 38010, Celaya, Gto., Mexico\\}

\author{J. A. Dunningham}
\email{J.Dunningham@sussex.ac.uk}
\address{Department of Physics and Astronomy, University of Sussex, Brighton, BN1 9QH, United Kingdom\\}
\begin{abstract}
\noindent
We model a gyroscope that exploits quantum effects in an atomic Bose-Einstein condensate to gain a tunable enhancement in precision. Current inertial navigation systems rely on the Sagnac effect using unentangled photons in fibre-optic systems and there are proposals for improving how the precision scales with the number of particles by using entanglement. 
Here we exploit a different route based on sharp resonances associated with quantum phase transitions. By adjusting the interaction between the particles and/or the shape of their trapping potential we are able to tune the width of the resonance and hence the precision of the measurement. Here we show how we can use this method to increase the overall sensitivity of a gyroscope by adjusting the system parameters as the measurement proceeds and our knowledge of the rotation improves. We illustrate this with an example where the precision is enhanced by a factor of more than 20 over the case without tuning, after 100 repetitions. Metrology schemes with tunable precision based on quantum phase transitions could offer an important complementary method to other quantum-enhanced measurement and sensing schemes. 

\end{abstract}
%
%03.65.-w Quantum mechanics
% Quantum mechanics, Q. Metrology, entanglement, 

\vspace{-1mm}
\maketitle
%
%
%%%%%%%%%%%%%%%%%%%%%%%%%%%%%%%%%%%%%
\section{Introduction~\label{sec1}}
%%%%%%%%%%%%%%%%%%%%%%%%%%%%%%%%%%%%%

%%%%%%% Initial opening section %%%%%%%%%%%%%%%%%%%%%%%%%%%%

Quantum metrology is expected to be one of the key enabling quantum technologies in the near future with great potential for advancing experimental science and industrial applications. Recent experiments have demonstrated quantum improvements in measuring protein concentration \cite{crespi2012measuring}, tracking lipid granules in yeast cells \cite{taylor2013biological}, searching for gravitational waves \cite{ligo2013enhancing}, and the detection of single-neuron activity \cite{barry2016optical}. The canonical approach to quantum metrology involves using entangled probe states to measure an external parameter with a precision that surpasses the standard quantum limit (SQL), where the uncertainty $\Delta\phi$ scales as $1/\sqrt{N}$, for $N$ particles. With carefully chosen states and measurements it is possible, at least in principle, to reach the Heisenberg limit (HL) where $\Delta\phi$ scales as $1/N$ \cite{Giovannetti2004}. However, in practice it is often better to look instead for advantages we can achieve in a given physical system with finite resources, rather than chase the HL. In this paper, we present a scheme in this spirit by considering what advantages can be achieved over classical schemes for specific gyroscopes rather than focussing on how the precision scales with the number of particles.

Recent work has shown that there is much more richness and subtlety to quantum metrology than the standard entanglement-enhanced approach suggests and it is possible, for example, to surpass the SQL by exploiting nonlinearities \cite{boixo2007generalized} or long-range order in condensed matter systems \cite{fernandez2017heisenberg} or without needing entanglement between interferometer paths \cite{knott2016local}, or between particles \cite{Higg, demkowicz2010multi}. Here we make use of quantum phase transitions and, while this involves entanglement, it does so in an indirect way that is different from what has previously been presented. We are more interested in the feature that quantum phase transitions have narrow resonances where the probe states rapidly vary as we scan over an external parameter, suggesting that they may be useful for precision metrology. This is backed up by the fact that the Fisher information, which is widely used to quantify measurement precisions, is proportional to the square of the rate at which the measurement-outcome probability distribution changes with the parameter of interest. However, these very narrow structures also have a major disadvantage since they can only measure the parameter over their very narrow range. In other words, we would need extremely good prior information about the value of the parameter we are measuring, which to some extent defeats the purpose of the measurement. In this work, we show how we can overcome this problem by proposing a scheme that exploits the narrow resonances but has a tunable precision. This means that we can start with a broad resonance matching our initial (poorly-defined) knowledge of the parameter and then, as measurements are made and information is obtained, the precision can be progressively improved by a kind of `boot-strapping' method. We demonstrate that this allows for significant advantages in measuring rotation rates which could, in turn, be important for the development of new inertial navigation devices as well as other wider applications.

{ Other work has investigated using phase transitions in quantum metrology to achieve the Heisenberg limit in different schemes \cite{Gammelmark, Tsang2013, Macieszczak, Rams, Porras}. Our approach focuses instead on tunable quantum-enhanced sensors and the advantages they can give with finite resources.} Although this idea may be able to be applied more generally, { such as in the coupling of a qubit with a laser at its critical point \cite{Porras} or the Jaynes-Cummings \cite{Mavrogordatos} and Dicke \cite{Rodrigues} models of atom-light coupled systems}, we focus here on one particular system for definiteness.
We consider trapped Bose-Einstein condensates (BECs) that are stirred and undergo a quantum phase transition to create a vortex. Unlike their classical counterparts, quantum phase transitions can be accessed even at zero temperature by varying physical parameters of the system. BECs are a promising system for future metrology experiments \cite{Fattori2008, Dunningham2002, Giovannetti2004, Ragole2016} since techniques have been developed that allow species with different properties to be condensed  \cite{cornish2000}, the trapping time to be increased, and new imaging techniques to be developed. Matter-wave interferometers using BECs or cold-atom clouds have already been used successfully in enhanced measurements of magnetic fields \cite{PhysRevLett.98.200801, PhysRevLett.113.103004}, gravitational fields \cite{peters1999measurement, 1367-2630-15-2-023009}  and rotations \cite{PhysRevLett.78.2046, PhysRevLett.97.240801, PhysRevA.80.061603}. 

{ The best available rotation sensors make use of fibre-optic and metre-scale ring laser gyroscopes \cite{Gebauer2020}. Atom interferometry offers a different approach with high stability and much smaller enclosed areas that are better suited to deployment in the field. Sagnac sensors, for example, use light fields to coherently spilt and recombine atomic wave packets in free fall and have dimensions on the centimetre scale \cite{Canuel06, Berg2015}. Another, even more compact, approach is to use waveguides or atomic traps. Other more-general schemes for ultraprecise measurements and sensing include systems such as hybrid atomic and optomechanical cavities \cite{Motazedifard2016,Motazedifard2021,Motazedifard2019}. }
Our scheme follows the compact atom-trap route and is based on the idea of Dagnino et~al. \cite{Dagnino2009} where a BEC is trapped in a quasi-two-dimensional weakly anisotropic potential and the rate of rotation of the potential is varied. As a BEC is unable to rotate at a non-quantised rate, it can only gain angular momentum by forming quantised vortices. Below a critical rotation rate of the potential, $\Omega_c$, the BEC will have zero angular momentum but, as we pass through the critical value, the BEC will undergo a quantum phase transition forming a quantised vortex. As this point, the ground state of the system experiences macroscopic symmetry-breaking accompanied by entanglement across the system and is dominated by two macroscopically occupied states. The phase transition is very sharp in the external rotation rate, which makes this system useful for metrology and, of particular interest is the fact that the width can be varied by changing other parameters such as the degree of anisotropy or the interaction strength between atoms. This feature allows us to propose a quantum-enhanced gyroscope with a tunable precision.

%%%%%%%%%%%%%%%%%%%%%%%%%%%%%%%%%%%%%%%%%%%%%%%%%%%%%%%%%%%%%%%
\section{MODEL\label{sec2}}
%%%%%%%%%%%%%%%%%%%%%%%%%%%%%%%%%%%%%%%%%%%%%%%%%%%%%%%%%%%%%%%

Our model consists of $N$ bosonic atoms of mass $M$ held in an axially symmetric harmonic potential with a frequency of $\omega_\perp$ in the $xy$ plane, and $\omega_z$ in the $z$ axis. Setting $\hbar\omega_z$ to be sufficiently large in comparison with the interaction energy forces all the bosons to occupy the lowest axial energy level, making the model behave quasi-two-dimensionally at low temperatures. For our calculations, we will take $N=6$, for computational convenience. Larger numbers of particles have been studied for this model in other work \cite{Rico-Gutierrez2013} and are qualitatively similar. We have all the features we need for the present scheme with $N=6$ and, since we are not concerned here with how the measurement precision scales with $N$, we do not need to consider more particles, which greatly simplifies our numerics. 
	
	The Hamiltonian in the rotating reference frame \cite{Rico-Gutierrez2015} is given by, 
	
	\begin{widetext}
	\begin{align}
	\label{Ham1}
	\hat{H} = \sum_{i = 1}^N \left( - \frac{\hbar^2}{2M} \nabla^2_i + \frac{1}{2}M \omega_z z_i^2 + \frac{1}{2}M \omega_\perp^2 \rho_i^2 +2AM\omega_\perp^2 (x_i^2 - y_i ^2) - \Omega L_{zi} \right) + \frac{1}{2} \sum_{j \neq k}^N \frac{g \hbar^2}{M} \delta (\bold{\overrightarrow{r_j}} - \bold{\overrightarrow{r_k}}) ,
	\end{align} 
	\end{widetext}
	
\noindent	
where $A$ is a dimensionless parameter that quantifies the $xy$ anisotropy in the potential and is taken to be small, $A \ll 1$, to ensure the solutions to the Hamiltonian converge \cite{Morris2006}; $\rho \equiv \sqrt{x^2 + y^2}$ is the radial coordinate in the $xy$ plane; $\Omega$ is the external rotation frequency; $L_{zi} $ is the $z$-component of the angular momentum of the $i$-th atom; and $M$ is the mass of an atom. The Hamiltonian is summed over the contributions from the $N$ atoms. The last term describes the energy of interactions between atoms and is quantified by the dimensionless parameter $g = a_s\sqrt{8\pi M\omega_z/\hbar}$, where $a_s$ is the 3D scattering length. For reasonable experimental parameters, $g$ takes values of order 0.1 and can be varied e.g. by changing $\omega_z$. { For a given value of $\omega_z$, the potential energy in the $z$-direction, i.e. $(M \omega_z z_i^2)/2$ can be ignored in the Hamiltonian because it is a constant since the system will always be in the lowest energy level in that direction for sufficiently large $\omega_z$.}

	If we scale the energies by $ \hbar \omega_\perp$, the second-quantised Hamiltonian \cite{LuisThesis} in the rotating frame is then, 
	
	\begin{widetext}
	\begin{equation}
	\label{Ham2}
	\hat{H} = 2 \sum_{k} n_k \hat{N}_k + \sum_{k} |m_k|\hat{N}_k - \Omega \hat{L} + \hat{N} + \sum_{k_1, k_2} V_{k_1 k_2}\hat{a}_{k_1}^\dagger \hat{a}_{k_2} + \frac{1}{2} \sum_{k_1, k_2} \sum_{l_1, l_2} U_{k_1 k_2 l_1 l_2} \hat{a}_{k_1} ^\dagger \hat{a}_{k_2}^\dagger \hat{a}_{l_1 }\hat{a}_{l_2} ,
	\end{equation}
	\end{widetext}
	
\noindent	
where $\hat{a}_k^\dagger$ creates a boson in the state $k = (n_k , m_k)$; $m_k$ specifies the number of units of angular momentum the particle has; $n_k$ specifies the Landau level; $\hat{N}_k=\hat{a}_k^\dagger \hat{a}$ is the occupation number operator for state $k$; and $\hat{L} = \sum_{k} m_k \hat{N}_k$ is the angular momentum operator for the system. 
The anisotropic term, $V_{k_1k_2}$ is given by,
\begin{equation}
\label{Anis}
V_{k_1k_2} = A \sqrt{\frac{n_{k_1} ! n_{k_2} !}{(n_{k_1} + |m_{k_1}|)!(n_{k_2} + |m_{k_2}|)!}} I_1(k_1,k_2)(\delta_{m_{k_2},m_{k_1 \pm 2}}),
\end{equation}
where
\begin{equation*}
\label{I1}
I_1(k_1,k_2) = \int_0^\infty e^{-x} x^\frac{|m_{k_1}| + |m_{k_2}| + 2}{2} L_{n_{k_1}}^{|m_{k_1}|}(x)L_{n_{k_2}}^{|m_{k_2}|}(x) dx.
\end{equation*}
The interaction $U_{k_1k_2l_1l_2}$ term is
\begin{equation}
\label{Int}
U_{k_1k_2l_1l_2} = \frac{g}{\pi} \frac{1}{2^\frac{\sum |m_t|}{2}} \sqrt{\Pi_t \frac{n_t !}{(n_t + |m_t|)!}} I_2 (k_1,k_2,l_1,l_2) (\delta_{m_{k_1} + m_{k_2}, m_{l_1} + m_{l_2}}),
\end{equation}
where
\begin{equation*}
\label{I2}
I_2(k_1,k_2,l_1,l_2) = \int_0 ^ \infty e^{-x} x^{\frac{\sum|m_t|}{2}}L_{n_{k_1}}^{|m_{k_1}|}(\frac{x}{2}) L_{n_{k_2}}^{|m_{k_2}|}(\frac{x}{2}) L_{n_{l_1}}^{|m_{l_1}|}(\frac{x}{2}) L_{n_{l_2}}^{|m_{l_2}|}(\frac{x}{2}) dx.
\end{equation*}

\noindent
$L_n^m(x)$ are Laguerre polynomials and $t$ takes values $k_1,k_2,l_1,l_2$ so that,
\begin{equation*}
\Pi_t \frac{n_t!}{(n_t + |m_t|)!} = \frac{n_{k_1}!n_{k_2}!n_{l_1}!n_{l_2}!}{(n_{k_1}+|m_{k_1|})!(n_{k_2}+|m_{k_2|})!(n_{l_1}+|m_{l_1|})!(n_{l_2}+|m_{l_2|})!}.
\end{equation*}

	As we might expect, the interaction term only couples states that have the same total angular momenta, which can be seen directly from the form of $U_{k_1k_2l_1l_2}$ in Eq.~(\ref{Int}). By contrast, the anisotropic part of the potential given by Eq.~(\ref{Anis}) connects states that have angular momenta that differ by $\pm 2$ units, which makes sense as the anisotropy is necessary to impart any angular momentum to the system. For our numerical simulations, we choose a Fock basis that, in the absence of any anisotropy, gives a block diagonal matrix in terms of angular momenta. The basis states are given by
	\begin{equation}
	 \ket{N_0,N_1,...} = \Pi_k \frac{\hat{a}_k^{\dagger^{N_k}}}{\sqrt{N_k} !} \ket{0}, \label{basiseq}
	 \end{equation}
where the $k$ index corresponds to both an angular momentum and a Landau level as specified above.

	In order to make the numerical calculation tractable, we need to truncate the basis and can do this consistently with the following two approximations. Firstly, we restrict the state to be in either the lowest Landau level or the first Landau level. The $n_{LL}$-th Landau level constraint is imposed by including only basis states which satisfy the relation, $1 + \left( \sum_{i=1}^N [n_i + (|m_i| - m_i)/2] \right) \leq n_{LL}$, where $n_{LL}$ is the Landau level upper constraint. For our calculations, which include the lowest and first excited Landau level, we set $n_{LL} = 2$. This condition means that a particle can have either one unit of angular momentum, one unit of radial excitation, or neither. We have numerically checked that adding another Landau level has a negligible effect on the critical frequency for the interaction strengths considered here as outlined in \cite{Rico-Gutierrez2013}, which justifies this truncation.

%\begin{figure}[t]
%  \centering
%    \includegraphics[width=0.5\textwidth]{Fig1.pdf}
%    %{Fig1.eps}
%    \caption{The quantum Fisher information, $F_q$ as a function of $\Omega$ for two different combinations of $g$ and $A$ values. The HWHM of the curves here are 0.0116$\omega_{\perp}$ and 0.0001$\omega_{\perp}$ respectively. The dashed horizontal lines show the maximum value the QFI can take for 6 unentangled particles ($F_q=6$). The particles are entangled in the regions where the curve lies above this line.}
%    \label{QFIfig}
%\end{figure}
	
	The second truncation follows the work of Morris and Feder \cite{Morris2006}, which shows that for a sufficiently small value of the anisotropy, states with a high angular momentum do not contribute to the many-body ground state. This allows us to restrict the basis to states with $ L \leq L_{max}$ where $L_{max}$ is taken as $L_{max} = N + 2$ and ensures convergence of the energies of the Hamiltonian \cite{Dagnino2009}. For a small non-zero value of $A$, the anisotropic term, Eq.~(\ref{Anis}), connects the subspaces separated by two units of total angular momentum.

As the rate of rotation of the condensate is adiabatically increased, the system changes from being at rest to containing one vortex, crossing a critical rotation rate, $\Omega_c$, that signals the symmetry-breaking quantum phase transition \cite{Dagnino2009}. In order to understand what is happening, it is convenient to describe the ground state in the region of $\Omega_c$ with a two mode (TM) approximation given by,

\begin{equation}
\ket{\Psi_{TM}} = \sum_{n=0}^{N/2} C_n \ket{N - 2n}_{\Psi_1} \ket{2n}_{\Psi_2},
\label{TM}
\end{equation}

\noindent
where $\ket{A}_{\Psi_1}\ket{B}_{\Psi_2}$ represents A(B) bosons in the most (second most) populated mode corresponding to state $\Psi_1 (\Psi_2)$. For clarity, this is not the same basis as given in Eq.~(\ref{basiseq}), but rather these two states are the eigenstates corresponding to the two largest eigenvalues of the single-particle density matrix (SPDM), $\rho_{kl} = \bra{\Psi}\hat{a_l}^\dagger \hat{a_k} \ket{\Psi} $. They have equal populations at $\Omega_c$ and account for almost all of the population, justifying the TM approximation. At very low rotations the most populated single-particle state $\Psi_1$ is almost entirely composed of $m = 0$ momentum particles. Just below $\Omega_c$,  $\Psi_1$ is composed of both $m = 0$ and $m = 2$ units of angular momentum \cite{Dagnino2009}. Just above $\Omega_c$, $\Psi_1$ and $\Psi_2$ switch occupancy with $\Psi_2$ becoming the most occupied state. This state is composed almost entirely of particles with $m = 1$ unit of momentum. By adjusting the system parameters, we can tune the width of the transition between these two cases. This is the physical process that is used for our measurement scheme. The TM approximation has been introduced here just to aid understanding of the processes involved and is not used in the simulation results presented in Sec.~\ref{sec3}.

\section{RESULTS\label{sec3}}
%%%%%%%%%%%%%%%%%%%%%%%%%%%%%%%%%%%%%%%%%%%%%%%%%%%%%%%%

\begin{figure}
  \centering
    \includegraphics[width=0.5\textwidth]{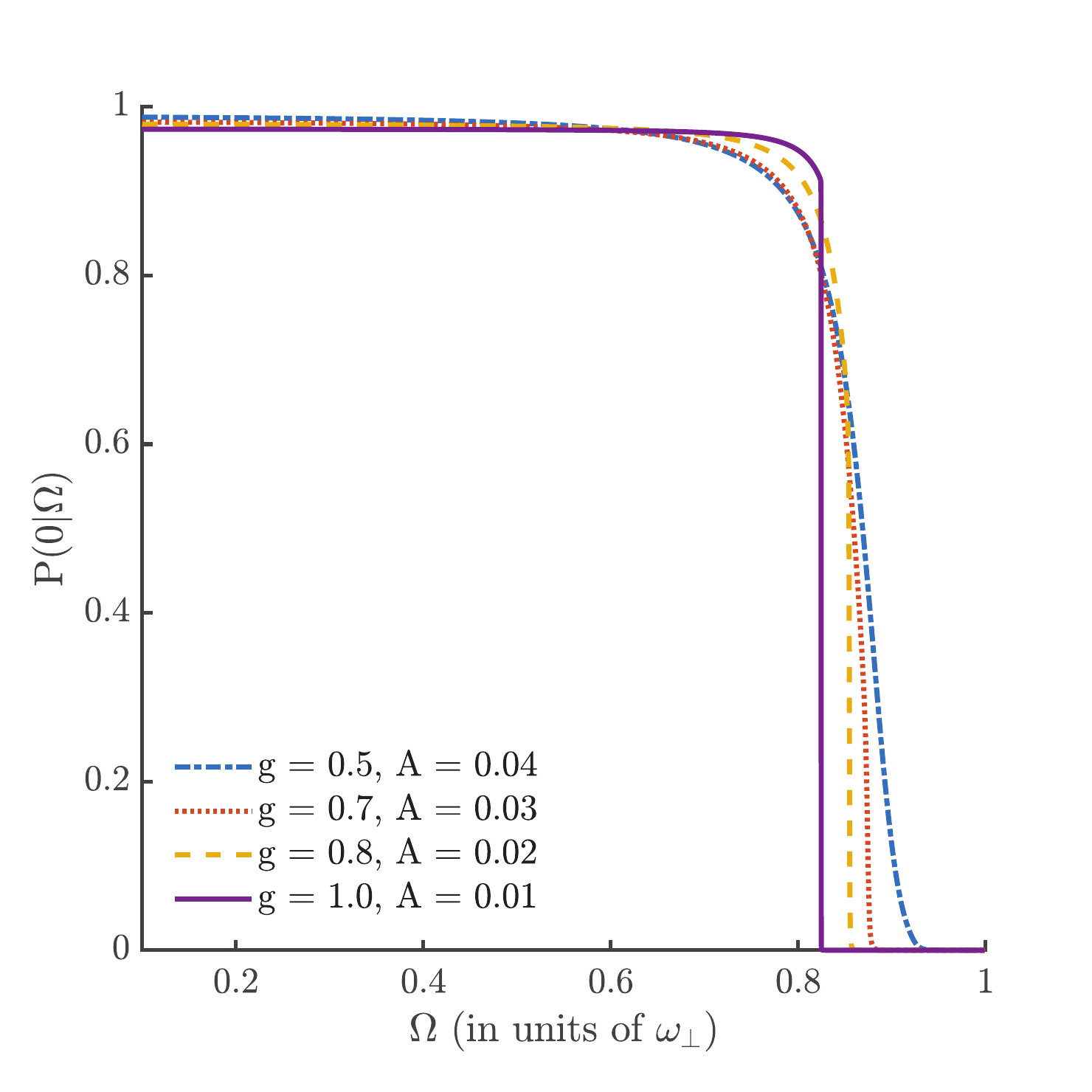}
    \caption{Plot of $P(0|\Omega)$ versus $\Omega$ for different combinations of $g$ and $A$ values. The gradient of $P(0|\Omega)$ can be varied by changing these parameters, enabling the precision of the scheme to be tuned.}
    \label{MonA0}
\end{figure}

We now describe the scheme that utilizes the results above to develop a quantum-enhanced gyroscope with tunable precision.
We use the fact that well below $\Omega_c$ all the particles have zero angular momentum and, as they pass through the phase transition, they acquire angular momentum. Our measurement protocol starts by calculating the probability that, for a given rotation rate, $\Omega$, all of the particles have zero angular momentum, $P(0|\Omega)$. A plot of this function for different values of $g$ and $A$ is shown in Fig.~\ref{MonA0}. For this calculation we use the full quantum state (with the two basis truncations discussed above) rather than the TM model to numerically find the ground state. Using the (truncated) basis given in Eq.~(\ref{basiseq}), the zero momentum states correspond to the kets with $k=(0,0)$ and $(0,1)$, i.e. states with zero angular momentum in either the lowest or first Landau levels.  {To determine the probability $P(0|\Omega)$, we numerically find the ground state of (\ref{Ham2}) and then sum the probabilities of the $k=(0,0)$ and $(0,1)$ components.}

The gradient of $P(0|\Omega)$ in Fig.~\ref{MonA0} determines the precision of the measurement scheme, but also limits its range. { Importantly, both these parameters can be adjusted by changing the values of $g$ and $A$ (as seen in Fig.~\ref{MonA0}), which allows us to tune the sensor.} We want to harness the power of the steep gradients without being constrained by their limited ranges. {It is not optimal to choose the transition region to be much broader than our prior knowledge of the rotation since, although that would ensure we can easily reach the transition region, the slope will not be as steep at it could have been. On the other hand, if the transition region is chosen to be much narrower than our prior knowledge, then we will often miss the transition region altogether since our best guess of the rotation is more uncertain than this. In such cases we will end up where $P(0|\Omega)$ has zero slope and will gain no information. The best choice, therefore, is when the transition region width is matched to our prior knowledge.} For our simulations, we take the prior to be a flat distribution over some range (though this is not necessary) and choose values of $g$ and $A$ to ensure that the region of steep gradient for $P(0|\Omega)$ covers the full range of the prior. For the values of $g$ and $A$ that we have investigated, this region can comfortably extend to $0.06 \omega_\perp$. Given a typical trapping frequency of 200 Hz, this means that the rotation must be known initially to within 12 Hz. This is well within the range required for applications such as inertial navigation.

The measurement scheme proceeds by adiabatically adding an external rotation to our system so that the sum of it and the mid-point of our prior information match the known $\Omega_c$ for our system. This guarantees that we will end up somewhere in the region of the steep gradient shown in Fig.~\ref{MonA0}, i.e. the region of high measurement sensitivity. The condensate is then rapidly (non-adiabatically) shifted out of the transition region to the lower side. This fixes the populations in the different levels, which can then be measured to calculate the most likely rotation rate.

\begin{figure}
	\includegraphics[width=0.5\textwidth]{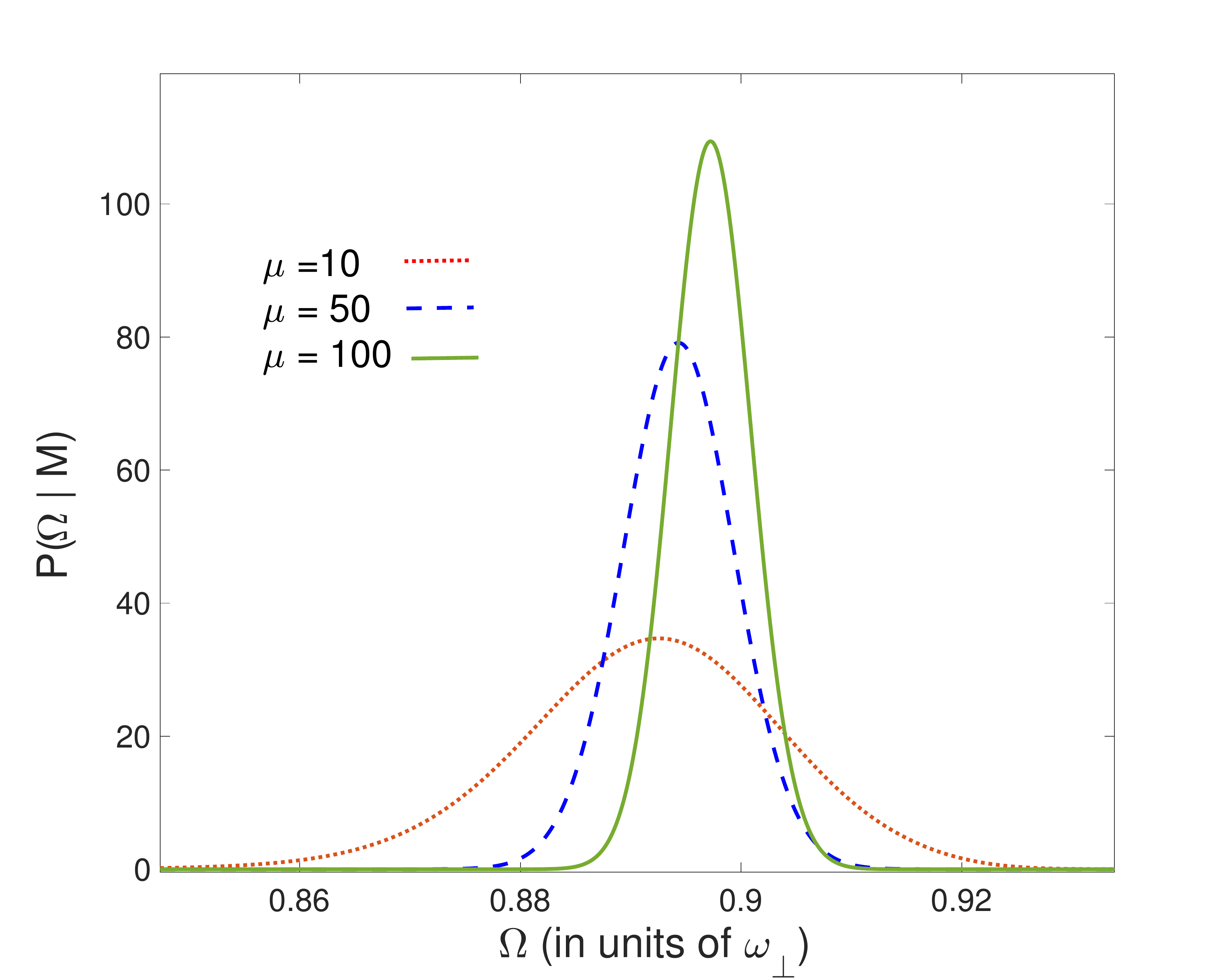}
	\caption{Bayesian prediction of $\Omega$, $P(\Omega|M)$,  shown for different numbers of measurements, $\mu$, where $M$ is a vector containing the $\mu$ measurement outcomes. As expected, the precision of the prediction improves as $\mu$ is increased. For the case shown here, $g=0.5$ and $A=0.04$ and the true rotation is $0.90\omega_{\perp}$. After 100 repeats the distribution was centered around 0.8980 with a standard deviation of $\sigma = 0.0037$; after 1000 repeats (not pictured due to scale), $\sigma$ was 0.0012; after $10^4$ repeats, $\sigma$ was $3.8 \times 10^{-4}$. These results have not yet made use of the tunability of the measurement scheme. }
	\label{fig:BayesNoOpt}
\end{figure}

The probability of all particles not rotating, $P(0|\Omega)$, is shown in Fig. \ref{MonA0} for different values of $g$ and $A$. Here we use the notation of $0$ to represent all particles not rotating and later $\neg 0$ to represent any other state. We generate a random number, $R$, to select an outcome from this distribution that models the outcome of a measurement. If $ R \leq P(0|\Omega)$ then we take all the particles to be not rotating and update the prior, using Bayesian methods, as $P(\Omega|0)\propto P(0|\Omega)P(\Omega)$. For any other measurement result, i.e. $R > P(0|\Omega)$, the prior is updated as $P(\Omega|\neg 0 ) \propto [1 - P(0|\Omega)]P(\Omega)$. Each iteration improves the measurement by incorporating the previous information, both honing in on the unknown rotation, Fig. \ref{fig:BayesNoOpt}, and increasing the precision by reducing the standard deviation, $\sigma$, of each curve.

	%------------------------------------------------

	\subsubsection*{Tuning the parameters} \label{Opt}

The results illustrated in Fig.~\ref{fig:BayesNoOpt} do not yet make use of the great advantage of this scheme, namely its tunability. In this section we show how, by tuning the system parameters as the measurement proceeds, it is possible to achieve smaller standard deviations with fewer measurements. 

We start by choosing values of $g$ and $A$ that give rise to a broad transition region as in Fig.~\ref{MonA0} with a width comparable to our (initially poor) prior knowledge of the rotation. The Bayesian estimation procedure described above is then run for a fixed number of repeats. This gives us improved knowledge of the rotation, given by a narrower Bayesian distribution as shown in Fig.~\ref{fig:BayesNoOpt}. We can now refine the next stage of the measurement by choosing a new combination of $g$ and $A$ values that produce a curve with a width that matches our updated Bayesian probability distribution. This process of updating the values of $g$ and $A$ to follow the width of the Bayesian curve is key to the tuning technique. In principle we could do this continuously, updating $g$ and $A$ after every measurement, but in practice it is simpler and still very effective to update the values only once or twice during a measurement run.

		\begin{figure}
		\includegraphics[width=0.5\linewidth]{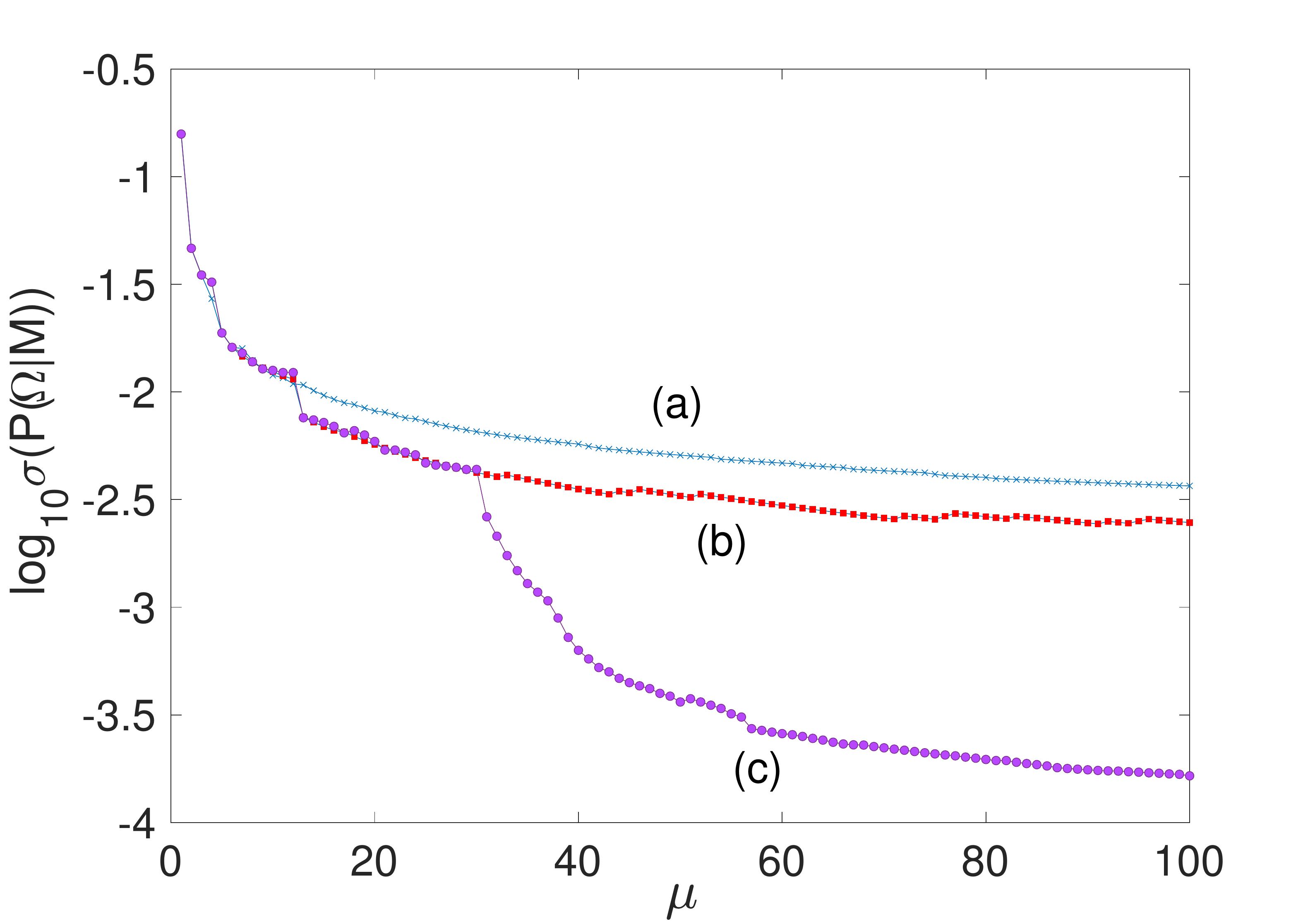}
		\caption{Logarithm of the standard deviation of the Bayesian distribution $P(\Omega |M)$ as a function of the number of measurements, $\mu$. Three different results are shown corresponding to (a) no tuning of the parameters, $g= 0.5$ and $A=0.04$, (b) one tuning after 12 measurements, and (c) two tunings after 12 and 32 measurements. The case with no tuning is the benchmark corresponding to the standard quantum limit (SQL). We see that tuning enables significant improvements in precision to be achieved.}
		\label{fig:BayesOpt}
	\end{figure}
	
The results of a simulation in Fig.~\ref{fig:BayesOpt} show how this technique can substantially improve a measurement. Line (a) uses only the initial values of $g=0.5$ and $A=0.04$ and serves as a reference benchmark against which we can compare our results. Line (b) starts with the same $g$ and $A$ as before but, after 10 measurements, these parameters are updated to match the width of the new Bayesian distribution. This gives an improvement in the standard deviation by a factor of about 1.5 after 100 measurements. Line (c) follows a similar procedure, but tunes the $g$ and $A$ values twice -- after 12 measurements and again after 32 measurements. In this case the precision of the measurement is improved by a factor of more than 20 after 100 measurements, when compared with the untuned case.  Alternatively, the untuned precision after $10^4$  measurements ($\sigma = 3.8 \times 10^{-4}$), can be achieved with fewer than 100 measurements using the tuning scheme shown in 5(c). {The large $\mu$ behaviour of each of the curves in Fig.~\ref{fig:BayesOpt} shows the $\mu^{-1/2}$ scaling that we would expect. Note that the `boot strapping' nature of this scheme is important. We would not get the same results just by initially choosing the final values of $g$ and $A$ in line (c). The reason for this is that the transition would then be much narrower than our prior knowledge of $\Omega$ and there would be a high probability that we would miss the transition altogether and gain no information from the measurement (see Fig.~\ref{MonA0}). The process of matching the width of the transition to our prior knowledge and then updating it as we gain more information is key.} The details of this scheme have been chosen just to illustrate the technique and larger gains could be achieved with an optimisation of the tuning strategy.

\section{Discussion}

The tunability of the phase transition has two major benefits. Firstly, we can start with a broad transition region, giving the experimentalist a large region to aim for when the rotation is not very well known. Secondly, we are not then bound by this rather imprecise measurement tool but can refine it as the measurement proceeds to gain further advantages in the overall precision. A nice feature of this scheme is that by varying either $g$ or $A$ we can achieve the desired tuning. This means that an experimentalist can choose to vary whichever is more practical.

	An efficient way of implementing this scheme could be to create an array of traps by using an optical lattice. By making use of the superfluid to Mott insulator phase transition it is possible to achieve an array where each of the traps contains the same fixed small number of atoms \cite{Greiner2002}, which is important for our scheme. It is also possible to achieve the Mott-like features that we need at finite temperatures \cite{Trotzky2010}. Once we are in the Mott regime, the optical lattice can be adiabatically transformed to create an array of traps with the desired populations and geometries. For our simulations, we model an array with 200 lattice sites, though this could be scaled up to larger numbers. By using this array, 200 results can be taken with a single measurement run. This is a big advantage for measuring time-dependent systems since, if the rotation rate is changing with time, we would need to gather sufficiently good statistics to measure it at a given time before it changes. By making all the measurements on a single shot, we greatly increase the bandwidth of our sensor. Our simulations use the initial values $g = 0.5$ and $A = 0.04$ and follow the procedure described above. A single run gives us 200 measurements in parallel and significantly changes the Bayesian distribution resulting in $\sigma = 0.0026$. At this point we apply the tuning and a second array is created and measured using updated values of $g$ and $A$.  These values are chosen as $g=0.6$ and $A=0.025$ to approximately match the width of the distribution, i.e. $\sigma = 0.0026$. The array method means that we can only alter the $g$ and $A$ values at measurement numbers corresponding to integer multiples of the number of traps. This restricts us a little but this restriction is far outweighed by the practical advantages of the scheme. After this process (i.e. 400 trap measurements in total) our simulation gives $\sigma = 1.8\times 10^{-4}$. This is a factor of 10 improvement over the untuned case with 400 measurements. It is possible to gain a bigger advantage with three or more tunings and the details of this can be worked out for specific implementations taking into account practical considerations.

%	\begin{figure}
%	\includegraphics[width=0.5\linewidth]{CombArrayEPS.eps}
%	\caption{Upper: The results of the initial unoptimised array of 200 trapping wells. The most likely rotation was found to be $0.852$, with $\sigma = 0.0102$. {\color{red} **why does this not mesh with the number in Fig.~4 - it is twice as many measurements so I would expect it to be ~$\sqrt{2}$ smaller. It seems to be either 3 times smaller or bigger, depending on where the decimal place should be in Fig.~4. Maybe it is because this is all for a different $g$ and $A$ than above and so is not directly comparable. This should be made clear.} \\
%		Lower: The results after the second array has been utilised, after one optimisation based on the results of the first array. The most likely position was found to be $0.8515$, with $\sigma = 0.00213$. This second array has values $g = 0.8$ and $A = 0.0015$ applied. As a reference, the vertical line marks the true position of the external rotation, $0.85$, as used in the simulation.}
%	\label{fig:BayesOptArray}
%\end{figure}    

	\begin{figure}
	\includegraphics[width=0.45\linewidth]{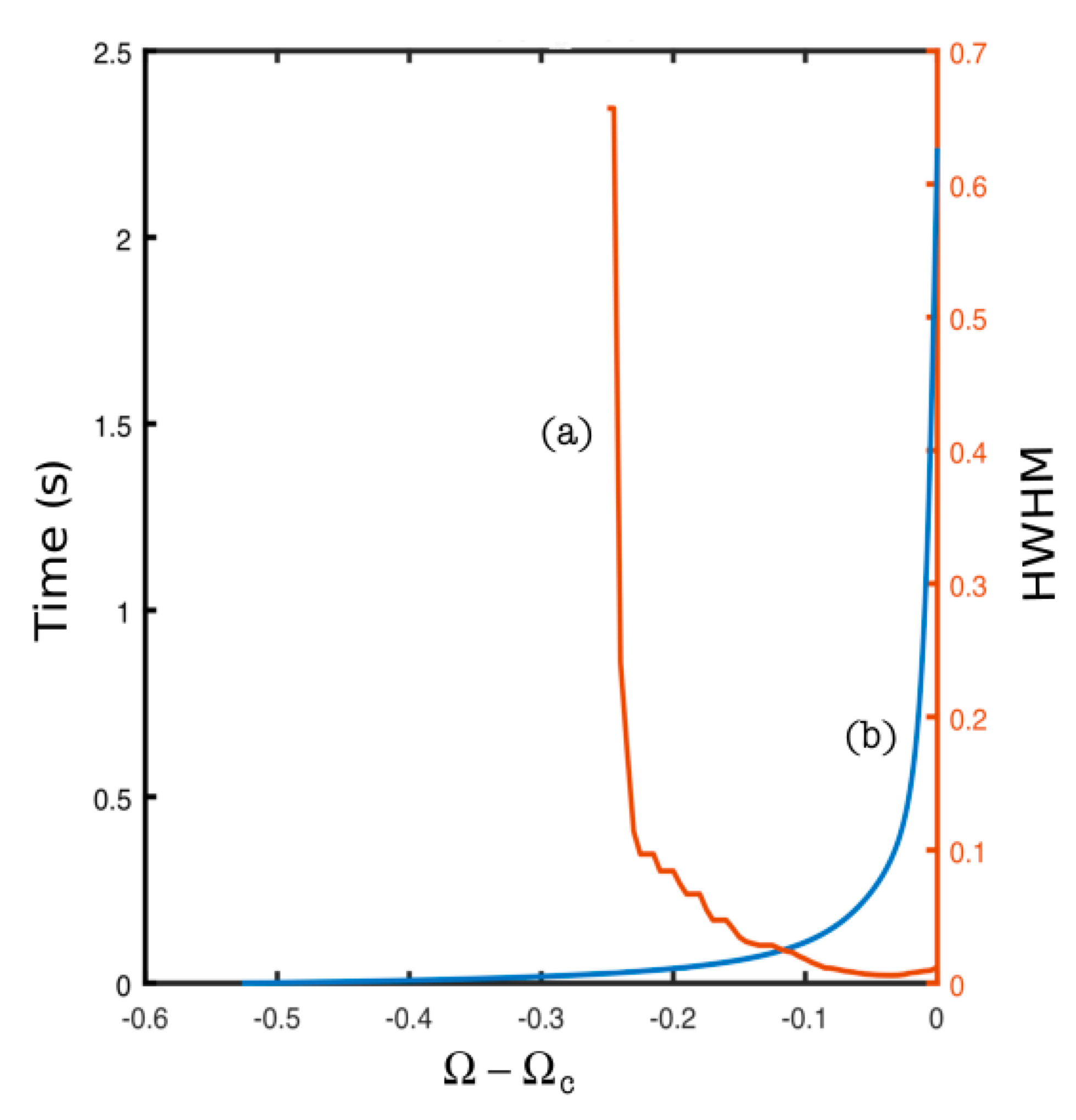}
	\caption{Effects of an offset in the rotation frequency, $\Omega$, from the critical frequency, $\Omega_c$ in the state preparation for the parameters $g=0.5$ and $A=0.04$. (a) Shows the final HWHM in units of $\omega_{\perp}$ of the 6-particle state (right hand scale) as a function of $\Omega - \Omega_c$ in units of $\omega_{\perp}$. (b) Shows the time in seconds (left hand scale) required to reach this offset frequency in order to maintain adiabaticity.}
	\label{fig:offset}
\end{figure} 

The read-out for this scheme only requires us to determine the probability of all the atoms having zero angular momentum. This is the same as the probability of being in the ground state away from the transition region. Once we we have performed the measurement process and rapidly shifted away from the transition region (as described above) we could remove all the excited states. One possible way of achieving this is by using Laguerre-Gaussian beams, which have a helical phase front and carry an orbital angular momentum \cite{Marzlin1997,Kapale2005,Vasnetsov2003}. The orbital angular momentum of the Laguerre-Gaussian beam can be selectively transferred to the centre of mass angular momentum mode of the particles that already have a non-zero angular momentum, exciting them out of the trap. An easier way may simply be to lower the trap height so that all but the ground state are not bound. Once the excited states have been removed, fluorescence spectroscopy could provide a readout mechanism telling us the fraction of traps in the ground state. We do not need to be able to count atoms, just the fraction of traps which still have atoms remaining in them. The practicalities for this are currently being investigated. 
 
 So far we have considered only pure states that are effectively at zero temperature. We should conclude by commenting on how noise is likely to affect our results. One of the strengths of our scheme is the simplicity of the read-out -- it is just a bucket detection where we only need to know the fraction of particles with zero angular momentum. {For a system with no classical noise sources such as finite temperature effects or laser fluctuations causing the trapping potential to jiggle, everything behaves as described above; if we do have these noise effects, we will get a mixed state that includes some component of the other levels. However, our read-out means that this does not affect things much. The lowest level acts just as before, but there is some depletion of the zero momentum state due to the noise.} Since we are only interested in how the fraction in the zero-momentum mode changes as a function of $\Omega$, we can get around the problem by first calibrating our system for $\Omega =0$. We are then able to measure other values of $\Omega$ by detecting changes in the ground state fraction using the method described above. 
 
 Although non-zero temperatures should not affect our scheme too much, the loss of particles is likely to be much more important. Many of the details of the scheme such as the energy levels and location of $\Omega_c$ depend on the fact that we have a particular number particles in each trap ($N=6$ for the case considered here). If particles are lost, this number changes and we get an error. We would avoid this problem if we could run the scheme much faster than the loss rate of particles. Trapped Bose condensates routinely have lifetimes of more than ten seconds \cite{Stellmer2009, Xiong2013} and so if our scheme was much quicker than this, loss would not be important. The problem is that our scheme relies on us remaining in the ground state by adiabatically changing the rotation of the trap. As we get close to the critical point, the energy gap gets very small and so the adiabatic process becomes slow. This puts a speed limit on how quickly we can change things and we need to compare this with the trap lifetime. 
 
It turns out that we can speed up the process and minimise the effects of loss by making use of the fact that we have a critical region (rather than a critical point) and so we do not need to go all the way to $\Omega_c$. This can be seen,  in Fig.~\ref{MonA0} where there is a clear range of $\Omega$ values over which enhanced measurements can be achieved. This means that a small distance from $\Omega_c$ (but still in the critical region) we can get a measurement advantage, but the adiabaticity criterion will not be so stringent and so the scheme can be run more quickly, reducing the effects of loss. A simulation of this for the case of $g=0.5$, $A=0.04$ is shown in Fig.~\ref{fig:offset}.  Line (b) shows the time needed to reach a reach a final rotation offset $(\Omega - \Omega_c)$. We see that, for zero offset, the time needed is a few seconds and so loss would play a role. The times get longer too for different values of $g$ and $A$ that give narrower resonances. For comparison, the precision (as measured by the HWHM) for each offset is plotted as line (a). There is generally a trade-off between the time taken and the precision achieved. However, we see that the HWHM does not change much over the range $\Omega - \Omega_c \in [-0.1\omega_{\perp}, 0]$ while the time needed to create this state varies by a factor of more than 10. This means that only going as far as $\Omega = \Omega_c - 0.1\omega_{\perp}$ in the state preparation will give a very similar measurement precision but take much less time. This not only allows us to develop schemes that are not affected by loss (by operating much faster than the trap lifetime), but is also highly desirable for creating higher bandwidth sensors, which are important for applications such as inertial navigation. 

{The adiabaticity criterion varies for different values of $g$ and $A$ and it is known that this  needs to be properly accounted for when considering scaling laws in metrology \cite{Rams}. In particular, the situation can change if we consider the precision we achieve as a function of experimental time rather than the number of measurements. We have checked Fig.~\ref{fig:BayesOpt} when plotted as a function of time and seen that the top two curves (i.e. no tuning and one tuning) largely overlap, meaning that there is no real advantage to the case we considered where the parameters were tuned once. By contrast, the case of two tunings (bottom curve) does still show a significant improvement, meaning that our scheme can still offer an advantage in the context of experimental time. However, this is something care needs to be taken over in any implementation.}

{To summarise, a proposed implementation would be as follows. Atoms are cooled into a Bose-Einstein condensate and then an optical lattice is turned on. By increasing the light intensity, a Mott insulator transition can occur where each lattice site has the same number of atoms. The lattice can then be deformed to ensure that the correct number of atoms, $N$, are in each trap of the array and that the $A$ and $g$ values are such that the width of the resonance matches our prior knowledge of the rotation rate $\Omega$. We then adiabatically increase the rotation rate of the array so that the middle of our prior knowledge plus the rotation we apply is in the critical transition region. We do not go all the way to the critical frequency, $\Omega_c$ to reduce the time needed, as discussed above. We then non-adiabatically reduce the rotation rate to bring the system out of the critical region. This locks the populations into the different levels. Finally, we reduce the trap heights so that all but the atoms in the zero-momentum states leave the traps and then we image the remaining population with fluorescence imaging techniques. This gives us $P(\Omega|M)$ and we can repeat the process with new values of $g$ and $A$ to match our updated knowledge of $\Omega$. }

{While we have demonstrated the principle of a tunable quantum-enhanced sensor, it is worth comparing its performance with existing matter wave gyroscopes to see if any advantage can be gained. Free-falling atom interferometers represent the state of the art in atomic inertial sensors and, in such a device, a sensitivity to rotation of $2.2\times 10^{-5}$ rad/s in 1s was achieved with $10^7$ atoms \cite{Canuel06}. Our scheme achieves $3.8\times 10^{-4}\omega_\perp$ with $10^4$ measurements, i.e. $6\times 10^4$ atoms. Extending to $~10^7$ atoms gives $~2.7\times 10^{-5}\omega_\perp$. The absolute performance depends on $\omega_\perp$ but since this is greater than 1, we are not yet at the level of  \cite{Canuel06}. However, our scheme has not been optimised, meaning bigger gains are likely through careful tuning of the parameters and reducing the value of $\omega_\perp$. This means that atomic lattice sensors could be competitive with matter-wave interferometers based on the Sagnac effect, but be much smaller with sub-millimeter dimensions compared with a few centimeters. Overall this shows that our scheme is not just an interesting new approach to quantum metrology, but has potential advantages too.}
 
 In conclusion, we have demonstrated a quantum-enhanced gyroscope scheme that, rather than relying on the enhanced number scaling that accompanies entangled particles, makes use of the sharp resonances associated with quantum phase transitions. The key feature of this scheme is that the precision can be tuned. This allows the measurement to be optimised through the course of a run and can lead to substantial gains in precision over unoptimised cases. {Our scheme has the potential to compete with existing schemes, but has additional size advantages.} We have also discussed possible ways of implementing this scheme and mitigating the effects of the main sources of error. Finally, while we have considered a very specific realisation in this paper, we believe that the idea of using quantum phase transitions for measurement schemes with tunable precision could be applicable in other systems \cite{Porras, Mavrogordatos, Rodrigues}. This may form an important new approach that complements more established ideas in quantum metrology and sensing.

\section*{Acknowledgments}
This work was partly funded by the UK Ministry of Defence through DSTL's National UK PhD programme~(contract number DSTLX-1000092185) and the United Kingdom EPSRC through the Quantum Technology Hub: Networked Quantum Information Technology (grant reference EP/M013243/1). We acknowledge helpful discussions with Luis Rico Guti\'errez.

%%%%%%%%%%%%%%%%%%%%%%%%%%%%%%%%%%%%%%%%%%%%%%%%%%%%%%%%%%%%%%%%%%%

%%%%%%%%%%%%%%%%%%%%%%%%%%%%%%%%%%%%%%
%%%%%%%%%%%%%%%%%%%%%%%%%%%%%%%%%%%%%%

%\bibliography{Bib1JabRef} 
%\bibliographystyle{unsrt}
\bibliographystyle{unsrt}

\end{document}